\documentclass[notitlepage]{revtex4-1}

\usepackage[utf8]{inputenc}
\usepackage[colorlinks=true,citecolor=blue,linkcolor=blue]{hyperref}
\usepackage[normalem]{ulem}
\usepackage{url}
\usepackage{graphicx,wrapfig,float,slashed,cancel}
\usepackage{amsmath,amssymb,epsfig,graphicx,xcolor,stmaryrd}
\usepackage{bm}
\usepackage{enumitem}
\usepackage{multirow}

\definecolor{darkblue}{RGB}{1, 90, 173}

\begin{document}


\title{Investigation of the  strange pentaquark candidate $P_{\psi s}^{\Lambda}(4338)^0$ recently observed by LHCb }

\author{K.~Azizi}
\email{ kazem.azizi@ut.ac.ir}
\thanks{Corresponding author}
\affiliation{Department of Physics, University of Tehran, North Karegar Avenue, Tehran
14395-547, Iran}
\affiliation{Department of Physics, Do\v{g}u\c{s} University, Dudullu-\"{U}mraniye, 34775
Istanbul, Turkey}
\affiliation{School of Particles and Accelerators, Institute for Research in Fundamental Sciences (IPM) P.O. Box 19395-5531, Tehran, Iran}
\author{Y.~Sarac}
\email{yasemin.sarac@atilim.edu.tr}
\affiliation{Electrical and Electronics Engineering Department,
Atilim University, 06836 Ankara, Turkey}
\author{H.~Sundu}
\email{ hayriyesundu.pamuk@medeniyet.edu.tr}
\affiliation{Department of Physics Engineering, Istanbul Medeniyet University, 34700 Istanbul, Turkey}

\date{\today}

\preprint{}

\begin{abstract}

The recently observed strange pentaquark candidate, $P_{\psi s}^{\Lambda}(4338){}^0$, is investigated to provide information about its nature and substructure. To this end, its mass and  width through the  decay channels $P_{\psi s}^{\Lambda}(4338){}^0 \rightarrow J/\psi \Lambda$ and $P_{\psi s}^{\Lambda}(4338){}^0 \rightarrow \eta_c \Lambda$ are calculated by applying two- and three-point QCD sum rules, respectively. The state is considered as a $\Xi_c\bar{D}$ meson-baryon molecular structure with spin-parity quantum numbers $J^P=\frac{1}{2}^-$. The obtained mass, $m_{P_{\psi s}^{\Lambda}(4338){}^0}=4338\pm 130~\mathrm{MeV}$, and width,  $\Gamma_{P_{\psi s}^{\Lambda}(4338){}^0}= 10.40\pm 1.93~\mathrm{MeV}$, are consistent with the experimental data within the presented uncertainties.  This allows us to  assign a  $\Xi_c\bar{D}$ molecular structure of $J^P=\frac{1}{2}^-$ for the $P_{\psi s}^{\Lambda}(4338){}^0$ state.

\end{abstract}


\maketitle

\renewcommand{\thefootnote}{\#\arabic{footnote}}
\setcounter{footnote}{0}
\section{\label{sec:level1}Introduction}\label{intro}

Exotic states such as pentaquarks and tetraquarks have become one of the focus of investigations in particle physics since the proposal of the quark model~\cite{Gell-Mann}. Because their existence was not prohibited either by the quark model or QCD,  they attracted attention from the beginning and were investigated extensively for a long time. Finally, expectations eventuated  and the announcement of the first observation of such states was made in 2003 for a tetraquark state, $X(3872)$, by the Belle Collaboration~\cite{Choi:2003ue}. Later, the confirmation of this state came from various collaborations~\cite{CDF:2003cab,BaBar:2004oro,D0:2004zmu,CDF:2009nxk,LHCb:2011zzp,CMS:2013fpt}. In 2015 a different member of the exotic states, namely the pentaquark state, containing five valance quarks was announced to be observed by the LHCb Collaboration~\cite{Aaij:2015tga}. The two states, $P_c(4380)$ and $P_c(4450)$, were observed in the $J/\psi+p$ decay channel~\cite{Aaij:2015tga} and later, in 2019, the analyses with a larger data sample revealed that the previously announced $P_c(4450)$ state split into two states, $P_c(4440)$ and $P_c(4454)$, and another pick, $P_c(4312)^+$, also came into sight~\cite{Aaij:2019vzc}. The reported resonance parameters for these states were as follows~ \cite{Aaij:2015tga,Aaij:2019vzc}: $ m_{P_c(4380)^+}=4380 \pm 8 \pm 29~\mbox{MeV}$, $\Gamma_{P_c(4380)^+}=205 \pm 18 \pm 86~\mbox{MeV}$, $ m_{P_c(4440)^+}=4440.3 \pm 1.3 ^{+ 4.1}_{-4.7}~\mbox{MeV}$, $\Gamma_{P_c(4440)^+}= 20.6 \pm 4.9^{+8.7}_{-10.1}~\mbox{MeV}$, $m_{P_c(4457)^+}=4457.3 \pm 0.6 ^{+ 4.1}_{-1.7}~\mbox{MeV}$, $\Gamma_{P_c(4457)^+}= 6.4 \pm 2.0^{+5.7}_{-1.9}~\mbox{MeV}$, $m_{P_c(4312)^+}=4311.9 \pm 0.7^{ +6.8}_{-0.6}~\mbox{MeV} $ and $\Gamma_{P_c(4312)^+}=9.8 \pm 2.7 ^{ +3.7}_{-4.5}~\mbox{MeV}$. In 2021 and 2022 there occurred two more pentaquark states' reports which possess a strange quark. These two states $P_{cs}(4459)$~\cite{LHCb:2020jpq} and $P_c(4337)$~\cite{LHCb:2021chn} were reported  to have the following masses and widths: $m_{P_{cs}(4459)^0}=4458.8 \pm 2.9^{+4.7}_{-1.1}~\mathrm{MeV}$, $\Gamma_{P_c(4459)^0}=17.3 \pm 6.5^{+8.0}_{-5.7}~\mathrm{MeV}$ and $m_{P_{cs}(4337)^+}=4337^{+7}_{-4}{}^{+2}_{-2}~\mathrm{MeV}$, $\Gamma_{P_c(4337)^+}= 29 ^{+26}_{-12}{}^{+14}_{-14}~\mathrm{MeV}$.

The experimental observations of these non-conventional states have increased the theoretical interest in these states and triggered extensive theoretical investigations over their identifications and various properties. Their substructures were still obscure, which has motivated many affords to explain this point by assigning them either being molecules or compact states. In Refs.~\cite{Wang:2015epa,Maiani:2015vwa,Anisovich:2015zqa,Li:2015gta,Lebed:2015tna,Anisovich:2015cia,Wang:2015ava,Wang:2015ixb,Ghosh:2015ksa,Wang:2015wsa,Wang:2016dzu,Zhang:2017mmw,Giannuzzi:2019esi,Wang:2019got,Wang:2020rdh,Ali:2020vee,Azizi:2021utt,Wang:2020eep,Azizi:2021pbh,Ozdem:2021ugy,Zhu:2015bba,Gao:2021hmv} the pentaquark states were investigated  by taking their substructure as diquark-diquark-antiquark or diquark-triquark forms. Owing to their proximity to the relevant meson baryon threshold and small widths, the molecular structure has been another commonly considered structure for the pentaquark states. With molecular structure assumption, the properties of these states, such as their mass spectrum and various interactions, were investigated with the application of different approaches including the contact-range effective field theory~\cite{Liu:2019tjn,Liu:2020hcv,Peng:2020gwk,PavonValderrama:2019nbk}, the effective Lagrangian approach~\cite{Cheng:2021gca,Ling:2021lmq,Xiao:2019mvs,Lu:2016nnt,Yang:2021pio}, the QCD sum rule method~\cite{Chen:2016otp,Azizi:2016dhy,Azizi:2018bdv,Azizi:2020ogm,Wang:2022neq,Wang:2022gfb,Wang:2022ltr,Wang:2021itn,Wang:2019hyc,Xu:2019zme}, one-boson exchange potential model ~\cite{Wang:2021hql,Chen:2022onm,Yalikun:2021dpk,Yalikun:2021bfm,Chen:2021kad,Wang:2021hql,Pan:2020xek,Liu:2019zvb,Wang:2019nwt,Chen:2019asm,Wang:2019aoc,Chen:2016ryt,Chen:2016heh} and quasipotential Bethe-Salpeter~\cite{He:2019ify,He:2019rva,Zhu:2021lhd,Wang:2022mxy}. Besides, one can find  other works in Refs.\cite{Chen:2015loa,He:2015cea,Chen:2015moa,Roca:2015dva,Meissner:2015mza,Xiao:2019gjd,Wang:2019nvm,Xiao:2019gjd,Chen:2020opr,Wu:2021caw,Yan:2021nio,Chen:2020uif,Du:2021fmf,Phumphan:2021tta,Lu:2021irg,Chen:2021obo,Peng:2022iez,Yang:2022ezl,Zhu:2022wpi,Wang:2022ztm,Giachino:2022pws,Ortega:2022uyu,Li:2023zag,Dong:2020nwk,Voloshin:2019aut,Peng:2020hql,Chen:2020kco,Chen:2021tip,Xiao:2021rgp,Du:2021bgb,Hu:2021nvs,Feijoo:2022rxf,Yalikun:2023waw,Nakamura:2022jpd} and the references therein  adopting the molecular interpretation for pentaquark states. They were also investigated with the possibility that they were arising from kinematical effects~\cite{Guo1,Mikhasenko:2015vca,Liu1,Bayar:2016ftu,Guo2,Nakamura:2021qvy}. Though there exist so many works over them, they were in need of many more to clarify or support their still uncertain properties. On the other hand, the possible pentaquark states other than the observed ones and possessing strange, bottom or charm quarks were also quested for with their expectation to be observed in the future~\cite{Liu:2022qns,Stancu:2022xax,Azizi:2022qll,Wang:2022ugk,Valera:2022elt,Giachino:2022pws,Hu:2022qlr,Wang:2022tib,Yan:2022wuz,Wang:2021hql,Wang:2015wsa,Azizi:2021pbh,Wang:2022neq,Shi:2021wyt,Deng:2022vkv,Wang:2020bjt,Azizi:2018dva,Meng:2019fan,Ozdem:2022iqk,Liu:2021ixf,Feijoo:2015kts,Chen:2015sxa,Cheng:2015cca,Xing:2021yid,Azizi:2017bgs,Huang:2021ave,Zhu:2020vto,Ferretti:2018ojb,Shimizu:2016rrd,Liu:2017xzo,Liu:2021efc,Liu:2021tpq,Yang:2022bfu,Xie:2020ckr,Zhang:2020dwp,Paryev:2020jkp,Xie:2020niw,Cao:2019gqo,Wang:2019zaw,Wang:2019ato,Huang:2018wed,Yang:2018oqd,Yamaguchi:2017zmn,Gutsche:2019mkg,Zhang:2020cdi,Ferretti:2020ewe,Wang:2021xao,An:2020jix,Yan:2021glh,Zhang:2020vpz}.

Among these pentaquark states, the present work focuses on the one which was observed very recently by the LHCb collaboration~\cite{LHCb:2022jad} in the amplitude analyses of $B^{-}\rightarrow J/\psi\bar{p}$ decay. The measured mass and width for the state, which was labeled as $P_{\psi s}^{\Lambda}(4338){}^0$, were reported as $m_{P_{\psi s}^{\Lambda}}=4338.2\pm 0.7 \pm 0.4~\mathrm{MeV}$ and $\Gamma_{P_{\psi s}^{\Lambda}} =7.0\pm 1.2\pm 1.3~\mathrm{MeV}$, respectively with the preferred spin-parity quantum numbers $J^P=\frac{1}{2}^{-}$. Having a mass and narrow width in consistency with meson-baryon molecular interpretation this structural form is adopted in Refs.~\cite{Ortega:2022uyu,Zhu:2022wpi}. In Ref.~\cite{Zhu:2022wpi}, a coupled-channel calculation was applied considering molecular states and the results obtained for $\Xi_c \bar{D}$ interaction indicated a wider peak than the observed one in the experiment. With the constituent quark model formalism $P_{\psi s}^{\Lambda}(4338){}^0$ was suggested to be a baryon-meson molecule state with $(I)J^P=(0)\frac{1}{2}^-$ and mass and width $m_{P_{\psi s}^{\Lambda}(4338){}^0}=4318.1~\mathrm{MeV}$ and $\Gamma_{P_{\psi s}^{\Lambda}(4338){}^0}=0.07~\mathrm{MeV}$, respectively~\cite{Ortega:2022uyu}. In the Ref.~\cite{Ozdem:2022kei}, the light cone QCD sum rules method is implemented to calculate the magnetic moments of the $P_{\psi s}^{\Lambda}(4338){}^0$ and $P_{\psi s}^{\Lambda}(4459){}^0$ states. The $P_{\psi s}^{\Lambda}(4338){}^0$ state and the other candidate pentaquark states were investigated in Ref.~\cite{Wang:2022neq} using QCD sum rules method and adopting the molecular structure, and the analyses were in favor of the $P_{\psi s}^{\Lambda}(4338){}^0$ state having a $\Xi_c\bar{D}$ molecular structure with spin-parity and isospin quantum numbers $J^P = \frac{1}{2}^-$ and $(I,I_3) = (0,0)$, respectively. The references~\cite{Karliner:2022erb,Yan:2022wuz,Meng:2022wgl,Chen:2022wkh} also investigated the $P_{\psi s}^{\Lambda}(4338){}^0$ state in association to the molecular form.

As already mentioned, there exist many studies devoted to describing the nature of the pentaquark states. These studies, performed with various approaches covering different structures for the pentaquark states, gave results consistent with the experimentally observed parameters. This fact makes the subject more intriguing and open to new investigations. Therefore it is necessary to provide further information to support or check the proposed alternative structures for a better identification of their obscure sub-structure. Moreover, the works over these exotic states both test our knowledge and provide support for the improvement of our understanding of the QCD in its non-perturbative regime. With these motivations, in the present work, we investigate two dominant strong decays of the recently observed pentaquark state $P_{\psi s}^{\Lambda}(4338){}^0$ into $ J/\psi \Lambda$ and $ \eta_c \Lambda$ states, sticking to a meson-baryon molecular interpretation. To this end, we apply the QCD sum rule method, which put forward its success with plenty of predictions consistent with the experimental observations~\cite{Shifman:1978bx,Shifman:1978by,Ioffe81}. The interpolating field for the state is chosen in the $\Xi_c \bar{D}$ molecular form. For completeness, firstly, we obtain the mass of the state and current coupling constant using the considered interpolating current, which are  subsequently to be used as inputs in strong coupling constant analyses. 

The rest of the paper has the following organization. In Sec.~\ref{II} the QCD sum rule for the mass of the considered state is presented with the numerical calculation of the corresponding results for the mass and current coupling constant. Sec.~\ref{III} contains the details of the QCD sum rules to calculate the strong coupling constants for the $P_{\psi s}^{\Lambda}(4338){}^0\rightarrow J/\psi \Lambda $ decay and their numerical analyses as well. Sec.~\ref{IV} presents the similar QCD sum rule calculation and the analyses for the strong coupling constants and the width corresponding to the $P_{\psi s}^{\Lambda}(4338){}^0\rightarrow \eta_c \Lambda $ channel.  Last section gives a short discussion and conclusion.

\section{QCD sum rule for the mass of $P_{\psi s}^{\Lambda}(4338){}^0$ state}\label{II}

To better understand the substructure of the pentaquark states, one way is the comparison of the observed properties of these particles with the related theoretical findings. One of the important observables  is the mass of these states. Beside the mass,  the current coupling constant is also a very important input that is needed to calculate the observables related to the decays of the particles like their width.  The present section gives the details of QCD sum rules calculations for the mass  and current coupling of the strange pentaquark candidate  $P_{\psi s}^{\Lambda}(4338){}^0$. The calculations start with the following two-point correlation function:
\begin{equation}
\Pi(q)=i\int d^{4}xe^{iq\cdot
x}\langle 0|\mathcal{T} \{J_{P_{cs}}(x)\bar{J}_{P_{cs}}(0)\}|0\rangle.
\label{eq:CorrFmassPc}
\end{equation}
In this equation,  $\mathcal{T}$ is the time ordering operator  and $J_{P_{cs}}$  represents the interpolating current for the $P_{\psi s}^{\Lambda}(4338){}^0$ pentaquark state, which is denoted as $P_{cs}$ in what follows. The current to interpolate this state is the  $\Xi_c\bar{D}$ molecular type with spin-parity  $J^P=\frac{1}{2}^-$ :
\begin{eqnarray}
J_{P_{cs}}&=&\epsilon^{abc}d^{T}_{a}C\gamma_5 s_b c_c \bar{c}_d i\gamma_5 u_{d},\label{CurrentPcs}
\end{eqnarray}
where $C$ represents the charge conjugation operator, subindices $a,~b,~c,~d$ are used to represent the color indices, and  $u,~d,~s,~c$ are the quark fields. To proceed in the calculations, one follows two separate paths resulting in two corresponding expressions containing the hadronic parameters on one side and QCD fundamental parameters on the other side. They are therefore called as the  hadronic and QCD sides, respectively. The physical parameter under quest is obtained via a match of these two sides by means of a dispersion relation. Both sides contain various Lorentz structures and the matching is carried out considering the same structures obtained in these representations. The Borel transformation and continuum subtraction are the final operations applied on both sides to suppress the contributions of higher states and continuum.

For the computation of the hadronic side, a complete set of the intermediate states with same quark content and carrying the same quantum numbers of the considered state is inserted inside the correlator. Treating the interpolating currents as annihilation or creation operators, and performing the integration over four-$x$ the correlator becomes
\begin{eqnarray}
\Pi^{\mathrm{Had}}(q)= \frac{\langle 0|J_{P_{cs}}|P_{cs}(q,s)\rangle \langle P_{cs}(q,s)|\bar{J}_{P_{cs}}|0\rangle}{m_{P_{cs}}^2-q^2}+\cdots,
\label{eq:masshadronicside1}
\end{eqnarray}
where the contributions coming from higher states and continuum are represented by $\cdots$, and one particle  pentaquark state with momentum $q$ and spin $s$ is represented by $|P_{cs}(q,s)\rangle$. To proceed, we need the following matrix element:
\begin{eqnarray}
\langle 0|\eta_{P_{cs}}|P_{cs}(q,s)\rangle &=& \lambda_{P_{cs}} u_{P_{cs}}(q,s).
\label{eq:matrixelement1}
\end{eqnarray}
given in terms of the Dirac spinor $u_{P_{cs}}(q,s)$ and current coupling constant $\lambda_{P_{cs}}$. Substituting  Eq.~(\ref{eq:matrixelement1}) into Eq.~(\ref{eq:masshadronicside1}) and applying the summation over spin 
\begin{eqnarray}
\sum_s u_{P_{cs}}(q,s)\bar{u}_{P_{cs}}(q,s)=\not\!q+m_{P_{cs}},
\end{eqnarray}
the result for the hadronic side is achieved as
\begin{eqnarray}
\Pi^{\mathrm{Had}}(q)=\frac{\lambda_{P_{cs}}^2(\not\!q+m_{P_{cs}})}{m_{P_{cs}}^2-q^2}+\cdots ,
\end{eqnarray}
which turns into following final form after the Borel transformation:
\begin{eqnarray}
\tilde{\Pi}^{\mathrm{Had}}(q)=\lambda_{P_{cs}}^2e^{-\frac{m_{P_{cs}}^2}{M^2}}(\not\!q+m_{P_{cs}})+\cdots,
\end{eqnarray}
where $\tilde{\Pi}^{\mathrm{Had}}(q)$ denotes the  Borel transformed form of the correlator and $M^2$ is the Borel mass parameter.

The QCD side of the calculations requires the usage of the interpolating field explicitly in the correlator, Eq.~(\ref{eq:CorrFmassPc}). This is followed by the possible contractions of the quark fields via Wick's theorem, which turns the result into the one containing quark propagators as
\begin{eqnarray}
\Pi^{\mathrm{QCD}}(q)&=&-i\int d^4x e^{iqx} \epsilon_{abc}\epsilon_{a'b'c'}\Big\{\mathrm{Tr}[S_s^{bb'}(x)\gamma_{5}CS_d^{T}{}^{aa'}(x)C\gamma_{5}]\mathrm{Tr}[S_u^{dd'}(x)\gamma_{5}S_c^{d'd}(-x)\gamma_{5}]\Big\} S_{c}^{cc'}(x).
\label{Eq:PiQCDmass}
\end{eqnarray}   
The light and heavy quark propagators necessary for further calculations have the following explicit forms~\cite{Yang:1993bp,Reinders:1984sr}:
\begin{eqnarray}
S_{q,}{}_{ab}(x)&=&i\delta _{ab}\frac{\slashed x}{2\pi ^{2}x^{4}}-\delta _{ab}%
\frac{m_{q}}{4\pi ^{2}x^{2}}-\delta _{ab}\frac{\langle \overline{q}q\rangle
}{12} +i\delta _{ab}\frac{\slashed xm_{q}\langle \overline{q}q\rangle }{48}%
-\delta _{ab}\frac{x^{2}}{192}\langle \overline{q}g_{\mathrm{s}}\sigma
Gq\rangle +i\delta _{ab}\frac{x^{2}\slashed xm_{q}}{1152}\langle \overline{q}%
g_{\mathrm{s}}\sigma Gq\rangle  \notag \\
&&-i\frac{g_{\mathrm{s}}G_{ab}^{\alpha \beta }}{32\pi ^{2}x^{2}}\left[ %
\slashed x{\sigma _{\alpha \beta }+\sigma _{\alpha \beta }}\slashed x\right]
-i\delta _{ab}\frac{x^{2}\slashed xg_{\mathrm{s}}^{2}\langle \overline{q}%
q\rangle ^{2}}{7776} ,  \label{Eq:qprop}
\end{eqnarray}%
and
\begin{eqnarray}
S_{c,{ab}}(x)&=&\frac{i}{(2\pi)^4}\int d^4k e^{-ik \cdot x} \left\{
\frac{\delta_{ab}}{\!\not\!{k}-m_c}
-\frac{g_sG^{\alpha\beta}_{ab}}{4}\frac{\sigma_{\alpha\beta}(\!\not\!{k}+m_c)+
(\!\not\!{k}+m_c)\sigma_{\alpha\beta}}{(k^2-m_c^2)^2}\right.\nonumber\\
&&\left.+\frac{\pi^2}{3} \langle \frac{\alpha_sGG}{\pi}\rangle
\delta_{ij}m_c \frac{k^2+m_c\!\not\!{k}}{(k^2-m_c^2)^4}
+\cdots\right\},
 \label{Eq:Qprop}
\end{eqnarray}
with  $G^{\alpha\beta}_{ab}=G^{\alpha\beta}_{A}t_{ab}^{A}$, $GG=G_{A}^{\alpha\beta}G_{A}^{\alpha\beta}$;  $a,~b=1,~2,~3$; $A=1,~2,\cdots,8$ and $t^A=\frac{\lambda^A}{2}$ where $\lambda^A$ are  the  Gell-Mann matrices. The propagator for $u$, $ d $ or $s$ quark  is represented by the sub-index $q$. The final results for this side are obtained after the Fourier and Borel transformations as
\begin{eqnarray}
\tilde{\Pi}_i^{\mathrm{QCD}}(s_0,M^2)=\int_{(2m_c+m_s)^2}^{s_0}dse^{-\frac{s}{M^2}}\rho_i(s)+\Gamma_i(M^2),
\label{Eq:Cor:QCD1}
\end{eqnarray}
where $s_0$ is the threshold parameter entering the calculations after the continuum subtraction application using the quark hadron duality assumption. $\rho_i(s)$ represents the spectral densities which are the imaginary parts of the results obtained as $\frac{1}{\pi}\mathrm{Im}\Pi_i^{\mathrm{QCD}}$ with $i$ corresponding to either the result obtained from the coefficient of the Lorentz structure $\not\!q$ or $I$. The results of such calculations contain long expressions and, to avoid giving overwhelming expressions in the text, the explicit results of spectral densities will not be presented here. The quantities that we seek in this section, namely mass and the current coupling constant of the pentaquark state, are obtained by the match of the coefficients of the same Lorentz structures obtained in both the hadronic and QCD sides. These matches are represented as
\begin{eqnarray}
\lambda_{P_{cs}}^2e^{-\frac{m_{P_{cs}}^2}{M^2}}&=&\tilde{\Pi}_{\not\!q}^{\mathrm{QCD}}(s_0,M^2),
\label{sumrulmasseq}
\end{eqnarray}
and
\begin{eqnarray}
\lambda_{P_{cs}}^2m_{P_{cs}}e^{-\frac{m_{P_{cs}}^2}{M^2}}&=&\tilde{\Pi}_{I}^{\mathrm{QCD}}(s_0,M^2).
\label{sumrulemassI}
\end{eqnarray}

\begin{figure}[h!]
\begin{center}
\includegraphics[totalheight=5cm,width=7cm]{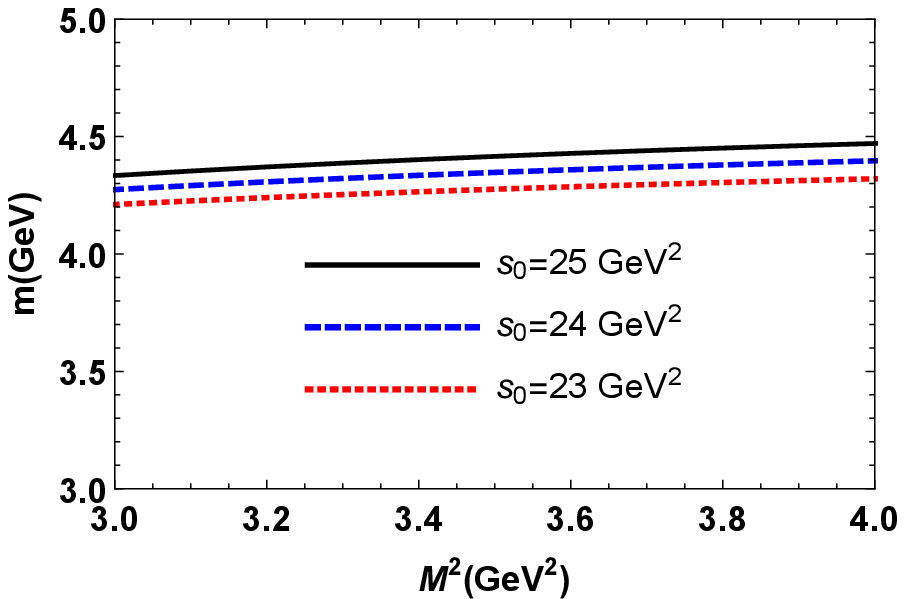}
\includegraphics[totalheight=5cm,width=7cm]{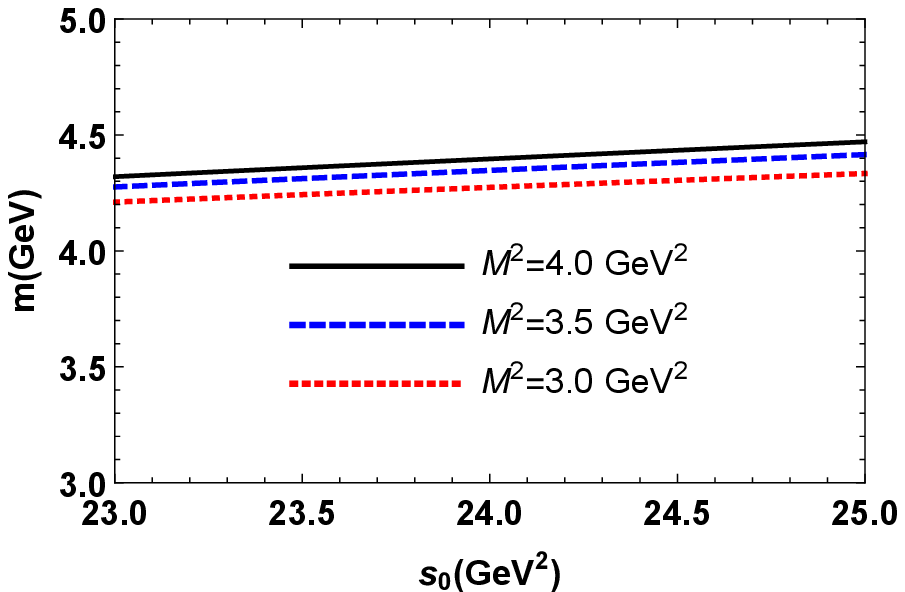}
\end{center}
\caption{\textbf{Left:} Variation of the the mass as function of $M^2$ at different values of threshold parameter $s_0$.
\textbf{Right:} Variation of the the mass as function of $s_0$ at different values of threshold parameter $M^2$. }
\label{gr:MassMsqS0}
\end{figure}

The next step is the analysis of the obtained results, for which one may apply any of the present structures. To this end, we choose the $I$ structure. The input parameters needed in the calculation of the mass and current coupling constant are given in Table~\ref{tab:Inputs}, which are also used for the coupling constant calculations to be given in the next section. 
\begin{table}[h!]
\begin{tabular}{|c|c|}
\hline\hline
Parameters & Values \\ \hline\hline
$m_{c}$                                     & $1.27\pm 0.02~\mathrm{GeV}$ \cite{Zyla:2020zbs}\\
$m_{s}$                                     & $93^{+11}_{-5}~\mathrm{MeV}$ \cite{Zyla:2020zbs}\\
$\langle \bar{q}q \rangle (1\mbox{GeV})$    & $(-0.24\pm 0.01)^3$ $\mathrm{GeV}^3$ \cite{Belyaev:1982sa}  \\
$\langle \bar{s}s \rangle $                 & $0.8\langle \bar{q}q \rangle$ \cite{Belyaev:1982sa} \\
$m_{0}^2 $                                  & $(0.8\pm0.1)$ $\mathrm{GeV}^2$ \cite{Belyaev:1982sa}\\
$\langle \overline{q}g_s\sigma Gq\rangle$   & $m_{0}^2\langle \bar{q}q \rangle$ \\
$\langle \frac{\alpha_s}{\pi} G^2 \rangle $ & $(0.012\pm0.004)$ $~\mathrm{GeV}^4 $\cite{Belyaev:1982cd}\\
$m_{J/\psi}$                                & $(3096.900\pm0.006)~\mathrm{MeV}$ \cite{Zyla:2020zbs}\\
$m_{\eta_c}$                                & $(2983.9\pm 0.4 )~\mathrm{MeV}$ \cite{Zyla:2020zbs}\\
$m_{\Lambda}$                             & $( 1115.683\pm 0.006 )~\mathrm{MeV}$ \cite{Zyla:2020zbs}\\
$\lambda_{\Lambda}$                       & $(0.013 \pm 0.02)~\mathrm{GeV}^3$ \cite{Aliev:2002ra}\\
$f_{J/\psi}$                                & $(481\pm36)~\mathrm{MeV}$ \cite{Veliev:2011kq}\\
$f_{\eta_c}$                                & $(320\pm 40)~\mathrm{MeV}$ \cite{Colangelo:1992cx}\\
\hline\hline
\end{tabular}%
\caption{Some input parameters used in the analyses of mass, current coupling constants and coupling constant of the $P_{cs}\rightarrow J/\psi \Lambda $ decay.}
\label{tab:Inputs}
\end{table} 
In addition to the given input parameters, there are two auxiliary parameters needed in the analyses:  the  Borel parameter $M^2$  and the continuum threshold $s_0$.  Following the standard criteria of the QCD sum rule method, their suitable intervals are fixed. These criteria include a relatively slight variation of the results with the change of these auxiliary parameters, the dominant contribution of the focused state compared to the higher states and continuum, and the convergence of the operator product expansion (OPE) used in the QCD side's calculation. Sticking to these criteria, we establish working regions of these parameters from the analyses. Seeking a region, for which the higher-order terms on OPE side contribute less compared to the lowest ones, and the ground  state dominates over the higher ones, the working interval of the Borel parameter is determined as:
\begin{eqnarray}
3.0~\mbox{GeV}^2\leq M^2\leq 4.0~\mbox{GeV}^2.
\end{eqnarray}
The determination of the continuum threshold interval has a connection to the energy of the possible excited states of the considered pentaquark state. With this issue in mind, we fix its interval as
\begin{eqnarray}
&23~\mbox{GeV}^2 \leq s_0 \leq 25~\mbox{GeV}^2.&
\end{eqnarray} 
By using all the inputs as well as the working windows of the auxiliary parameters,
we depict the variation  of the  mass with respect to  the auxiliary parameters for the considered structure  in  Fig.~\ref{gr:MassMsqS0}. 
This figure shows the mild dependence of the mass  on  the variations of the  auxiliary parameters in their working windows.  The residual dependence appear as the uncertainties in the results.

The resultant values for the mass and the current coupling constant are:  
\begin{eqnarray}
m_{P_{cs}}=4338\pm 130~\mathrm{MeV},~~~~~~\mathrm{and}~~~~~~\lambda_{P_{cs}}=(7.24\pm 0.21)\times10^{-4}~\mathrm{GeV}^6.
\end{eqnarray}
The result obtained for the mass has a good consistency with the observed mass of $P_{\psi s}^{\Lambda}(4338){}^0$ state announced as $m_{P_{\psi s}^{\Lambda}}=4338.2\pm 0.7 \pm 0.4~\mathrm{MeV}$ ~\cite{LHCb:2022jad}. 

As is mentioned,  the results obtained in this section are necessary inputs for the next sections which are devoted to the strong decays of the considered pentaquark state, namely $P_{\psi s}^{\Lambda}(4338){}^0\rightarrow J/\psi \Lambda $ and $P_{\psi s}^{\Lambda}(4338){}^0\rightarrow \eta_c \Lambda $.

\section{QCD sum rule to analyze the $P_{\psi s}^{\Lambda}(4338){}^0\rightarrow J/\psi \Lambda $ decay}\label{III}

The bare mass investigations of pentaquark states present in the literature, performed to explain the properties of the newly observed states, indicated that different assumptions for the substructures of these states might give consistent mass predictions with the observed ones. These necessitate deeper investigations which serve as support for previous findings. With this motivation, to clarify more the substructure and the quantum numbers of the observed $P_{\psi s}^{\Lambda}(4338){}^0$ state, in this section, we investigate the  $P_{\psi s}^{\Lambda}(4338){}^0\rightarrow J/\psi \Lambda $ decay and calculate its width.  To this end, the main ingredients are the strong coupling constants entering the low energy amplitude of the decay. To calculate these coupling constants via the QCD sum rule method, we use the following three-point correlation function:
\begin{equation}
\Pi_{\mu} (p, q)=i^2\int d^{4}xe^{-ip\cdot
x}\int d^{4}ye^{ip'\cdot y}\langle 0|\mathcal{T} \{J^{\Lambda}(y)
J_{\mu}^{J/\psi}(0)\bar{J}^{P_{cs}}(x)\}|0\rangle,
\label{eq:CorrF1Pc}
\end{equation}
with the interpolating currents given in Eq.~(\ref{CurrentPcs}) and
\begin{eqnarray}
J^{\Lambda}&=&\frac{1}{\sqrt{6}}\epsilon^{lmn}\sum_{i=1}^{2}\Big[2(u^{T}_l CA_1^i d_m)A_2^i s_n+(u^{T}_l CA_1^i s_m)A_2^i d_n+(d^{T}_n CA_1^i s_m)A_2^i u_l\Big],\nonumber\\
J_{\mu}^{J/\psi}&=&\bar{c}_l\gamma_{\mu}c_l.
\label{InterpFields}
\end{eqnarray}
In Eq.~(\ref{InterpFields}) sub-indices, $l,~m,~n$, are used to represent the color indices, and $u,~s,~c$ stand for quark fields, $A_1^1=I$, $A_1^2=A_2^1=\gamma_5$, $A_2^2=\beta$ which is a mixing parameter, and $C$ represents the charge conjugation operator. Similar steps of the calculation followed in the previous section also apply here. Calculation of hadronic and QCD sides are followed by their proper matches considering the coefficients of the same Lorentz structures from both sides.

For the hadronic side, we insert complete sets of hadronic states that have the same quantum numbers with the interpolating fields. Taking the four integral results in
\begin{eqnarray}
\Pi _{\mu }^{\mathrm{Had}}(p, q)=\frac{\langle 0|J^{\Lambda }|\Lambda(p',s')\rangle \langle 0|J_{\mu }^{J/\psi}|J/\psi(q)\rangle \langle J/\psi(q) \Lambda(p',s')|P_{cs}(p,s)\rangle \langle P_{cs}(p,s)|\bar{J}^{P_{cs}}|0\rangle }{(m_{\Lambda}^2-p'^2)(m_{J/\psi}^2-q^2)(m_{P_{css}}^2-p^2)}+\cdots,  \label{eq:CorrF1Phys}
\end{eqnarray}
with $\cdots$ denoting the contribution of the higher states and continuum; and $p$, $p'$ and $q$ being the respective momenta of the  $P_{cs}$,  $\Lambda$ and $J/\psi$ states. The required matrix elements for calculations have the following forms:
\begin{eqnarray}
\langle 0|J^{P_{cs}}|P_{cs}(p,s)\rangle &=&\lambda_{P_{cs}} u_{P_{cs}}(p,s),
\nonumber\\
\langle 0|J^{\Lambda }|\Lambda(p',s')\rangle &=&\lambda_{\Lambda} u_{\Lambda}(p',s'),
\nonumber\\
\langle 0|J_{\mu }^{J/\psi}|J/\psi(q)\rangle &=&f_{J/\psi} m_{J/\psi} \varepsilon_{\mu},
\label{eq:ResPcss}
\end{eqnarray}
where $ \varepsilon_{\mu} $ and  $ f_{J/\psi} $ represent the polarization vector and the decay constant of the $J/\psi$ state ; and $\lambda_{P_{cs}}$ and $\lambda_{\Lambda}$ are the current coupling constants of the $P_{cs}$ and $\Lambda$ states, respectively. $|P_{cs}(p,s)\rangle$ corresponds to the one-particle pentaquark state with its spinor $ u_{P_{cs}}$ and  $ u_{\Lambda}$  is the spinor of $\Lambda$ state. The matrix element,$ \langle J/\psi(q) \Lambda(p',s')|P_{cs}(p,s)\rangle$  is given in terms of the considered strong coupling constants, $g_1$ and $g_2$, as
\begin{eqnarray}
\langle J/\psi(q) \Lambda(p',s')|P_{cs}(p,s)\rangle = \varepsilon^{* \mu}\bar{u}_{\Lambda}(p',s')\big[g_1\gamma_{\mu}-\frac{i\sigma_{\mu\alpha}}{m_{\Lambda}+m_{P_{cs}}}q^{\alpha}g_2\big]\gamma_5 u_{P_{cs}}(p,s).
\label{eq:coupling}
\end{eqnarray}
Substituting the matrix elements, Eq.~(\ref{eq:ResPcss}) and Eq.~(\ref{eq:coupling}), into the Eq.~(\ref{eq:CorrF1Phys}) using the summation over the spins of the spinors and polarization vector given as
\begin{eqnarray}
\sum_{s}u_{P_{cs}}(p,s)\bar{u}_{P_{cs}}(p,s)&=&({\slashed
p}+m_{P_{cs}}),\nonumber \\
\sum_{s'}u_{\Lambda}(p',s')\bar{u}_{\Lambda}(p',s')&=&({\slashed
p'}+m_{\Lambda}), \nonumber\\
\varepsilon_{\alpha}\varepsilon^*_{\beta}&=&-g_{\alpha\beta}+\frac{q_\alpha q_\beta}{m_{J/\psi}^2},
\label{eq:SumPc}
\end{eqnarray}
the hadronic side is achieved as
\begin{eqnarray}
\tilde{\Pi}_{\mu }^{\mathrm{Had}}(p, q)&=&e^{-\frac{m_{P_{cs}^2}}{M^2}}e^{-\frac{m_{\Lambda}^2}{M'^2}}\frac{f_{J/\psi} \lambda_{\Lambda} \lambda_{P_{cs}}  m_{\Lambda} }{ m_{J/\psi} (m_{\Lambda} + m_{P_{cs}}) (m_{J/\psi}^2 + Q^2)}\big[- g_1 (m_{\Lambda} + m_{P_{cs}})^2+g_2 m_{J/\psi}^2 \big] \not\!p p_{\mu} \gamma_5
 \nonumber\\
&+&e^{-\frac{m_{P_{cs}^2}}{M^2}}e^{-\frac{m_{\Lambda}^2}{M'^2}}\frac{f_{J/\psi} \lambda_{\Lambda} \lambda_{P_{cs}}m_{J/\psi} m_{\Lambda}  }{  (m_{\Lambda} + m_{P_{cs}}) (m_{J/\psi}^2 + Q^2)}\big[ g_1 (m_{\Lambda} + m_{P_{cs}})+g_2 (m_{\Lambda} - m_{P_{cs}}) \big] \not\!p \gamma_{\mu} \gamma_5 \nonumber\\
&+&
\mathrm{other~structures}+\cdots.
\label{eq:haddecay}
\end{eqnarray}
Among the present Lorentz structures the ones used in the analyses are given explicitly, and the remaining ones are represented by $\mathrm{other~structures}$. Here  $Q^2=-q^2$. The Borel parameters  $M^2$ and $M'^2$, present in the last result,  are determined from the analyses following similar criteria given in the previous section.

As for the QCD side, the insertion of the interpolating currents given in Eqs.~(\ref{CurrentPcs}) and (\ref{InterpFields}) inside the correlator in  Eq.~(\ref{eq:CorrF1Pc}),  and after the possible contractions of quark fields using the Wick's theorem, the result takes the following form in terms of the quark propagators:
\begin{eqnarray}
\Pi_{\mu} ^\mathrm{QCD}(p, q)&=&i^2\int d^{4}xe^{-ip\cdot
x}\int d^{4}ye^{ip'\cdot y}\frac{1}{\sqrt{6}} \epsilon^{abc} \epsilon^{a'b'c'}\sum_{i=1}^{2}\Big[\mathrm{Tr}[S_s^{bb'}(y-x)\gamma_5CS_d^T{}^{ca'}(y-x)CA_1^i]A_2^iS_u^{ad'}(y-x)\nonumber\\
 &\times &\gamma_5S_c^{d'l}(x)\gamma_{\mu}S_c^{lc'}(-x)-2A_2^iS_s^{cb'}(y-x)\gamma_5 C S_d^T{}^{ba'}(y-x)CA_1^iS_u^{ad'}(y-x)\gamma_5S_c^{d'l}(x)\gamma_{\mu}S_c^{lc'}(-x)\nonumber\\
 &-& A_2^iS_d^{ca'}(y-x)\gamma_5 C S_s^T{}^{bb'}(y-x)CA_1^iS_u^{ad'}(y-x)\gamma_5S_c^{d'l}(x)\gamma_{\mu}S_c^{lc'}(-x)\Big].
\label{eq:CorrF1QCD}
\end{eqnarray}
Considering the same Lorentz structures given explicitly in  Eq.~(\ref{eq:haddecay}), we obtain the QCD side and represent the lengthy results shortly as in the following form:
\begin{eqnarray}
\Pi_{\mu }^{\mathrm{QCD}}(p,q)&=&\Pi_1\,\not\!p p_{\mu} \gamma_5 +
\Pi_2\, \not\!p \gamma_{\mu} \gamma_5 +\mathrm{other \,\,\, structures}. \label{eq:PiOPE}
\end{eqnarray}
 To proceed in the calculations, the light and heavy quark propagators given in Eqs.~(\ref{Eq:qprop}) and (\ref{Eq:Qprop}) are used explicitly and the four-dimensional Fourier integrals are performed. The imaginary parts of the obtained results constitute the spectral densities to be used in the following relation:
\begin{eqnarray}
\Pi_{i}=\int ds\int
ds'\frac{\rho_{i}^{\mathrm{pert}}(s,s',q^{2})+\rho_{i}^{\mathrm{non-pert}}(s,s',q^{2})}{(s-p^{2})
(s'-p'^{2})},  \label{eq:Pispect}
\end{eqnarray}
where $\rho_i(s,s',q^2)=\frac{1}{\pi}\mathrm{Im}[\Pi_i]$; and $ \rho_{i}^{\mathrm{pert}}(s,s',q^{2}) $ and $ \rho_{i}^{\mathrm{non-pert}}(s,s',q^{2}) $ represent the results of the perturbative and non-perturbative parts, respectively, with $i=1,~2,..,12$ corresponding to  all the Lorentz structures existing  in the results. The analyses in the present work are performed via resulting matches of the hadronic and QCD sides obtained from the structures $i=1,~2$, which results in two coupled sum rules equations including both $g_1$ and $g_2$: 
\begin{eqnarray}
&&e^{-\frac{m_{P_{cs}^2}}{M^2}}e^{-\frac{m_{\Lambda}^2}{M'^2}}\frac{f_{J/\psi} \lambda_{\Lambda} \lambda_{P_{cs}}  m_{\Lambda} }{ m_{J/\psi} (m_{\Lambda} + m_{P_{cs}}) (m_{J/\psi}^2 + Q^2)}\big[- g_1 (m_{\Lambda} + m_{P_{cs}})^2+g_2 m_{J/\psi}^2 \big] =
\tilde{\Pi}_1,
\\
&&e^{-\frac{m_{P_{cs}^2}}{M^2}}e^{-\frac{m_{\Lambda}^2}{M'^2}}\frac{f_{J/\psi} \lambda_{\Lambda} \lambda_{P_{cs}}m_{J/\psi} m_{\Lambda}  }{  (m_{\Lambda} + m_{P_{cs}}) (m_{J/\psi}^2 + Q^2)}\big[ g_1 (m_{\Lambda} + m_{P_{cs}})+g_2 (m_{\Lambda} - m_{P_{cs}}) \big]  =
\tilde{\Pi}_2,
\label{eq:SRg1g2}
\end{eqnarray}
where the Borel transformations on the variables $-p'^2$ and $-p^2$ have been  performed, and $\tilde{\Pi}_i$ in the results represent the Borel transformed $\Pi_i$ expressions obtained in the QCD side. Solution of these equations for $g_1$ and $g_2$ give
\begin{eqnarray}
g_1&=&
e^{\frac{m_{P_{cs}}^2}{M^2}}e^{\frac{m_{\Lambda}^2}{M'^2}}\frac{m_{J/\psi}(m_{J/\psi}^2+Q^2)
\left[(m_{P_{cs}}-m_{\Lambda})\tilde{\Pi}_1 +\tilde{\Pi}_2 \right]}
{f_{J/\psi}\lambda_{\Lambda}\lambda_{P_{cs}}m_{\Lambda}(m_{\Lambda}^2+m_{J/\psi}^2-m_{P_{cs}}^2)},\nonumber \\
g_2&=& e^{\frac{m_{P_{cs}}^2}{M^2}}e^{\frac{m_{\Lambda}^2}{M'^2}}\frac{(m_{P_{cs}}+m_{\Lambda})(m_{J/\psi}^2+Q^2)
\left[m_{J/\psi}^2\tilde{\Pi}_1+(m_{P_{cs}}+m_{\Lambda})\tilde{\Pi}_2  \right]}
{f_{J/\psi}\lambda_{\Lambda}\lambda_{P_{cs}}m_{\Lambda}m_{J/\psi}(m_{\Lambda}^2+m_{J/\psi}^2-m_{P_{cs}}^2)}.
  \label{eq:g1g2}
\end{eqnarray}

\begin{figure}[h!]
\begin{center}
\includegraphics[totalheight=5cm,width=7cm]{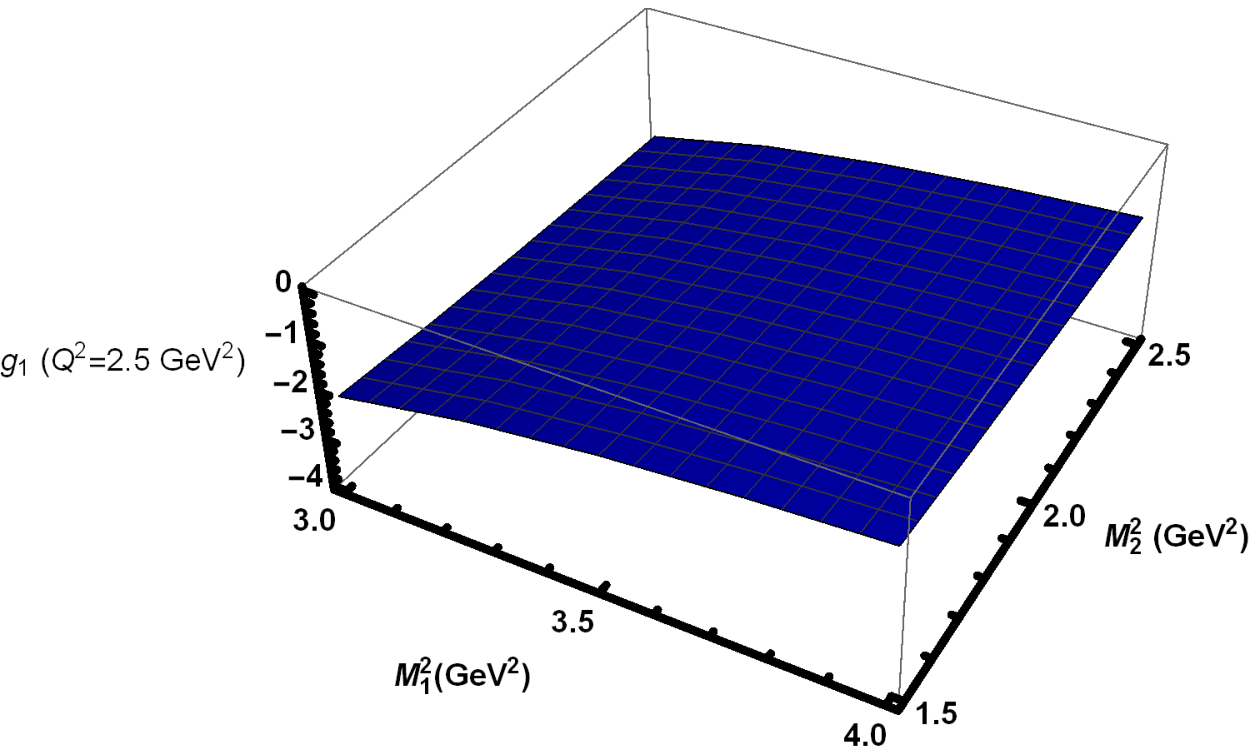}
\includegraphics[totalheight=5cm,width=7cm]{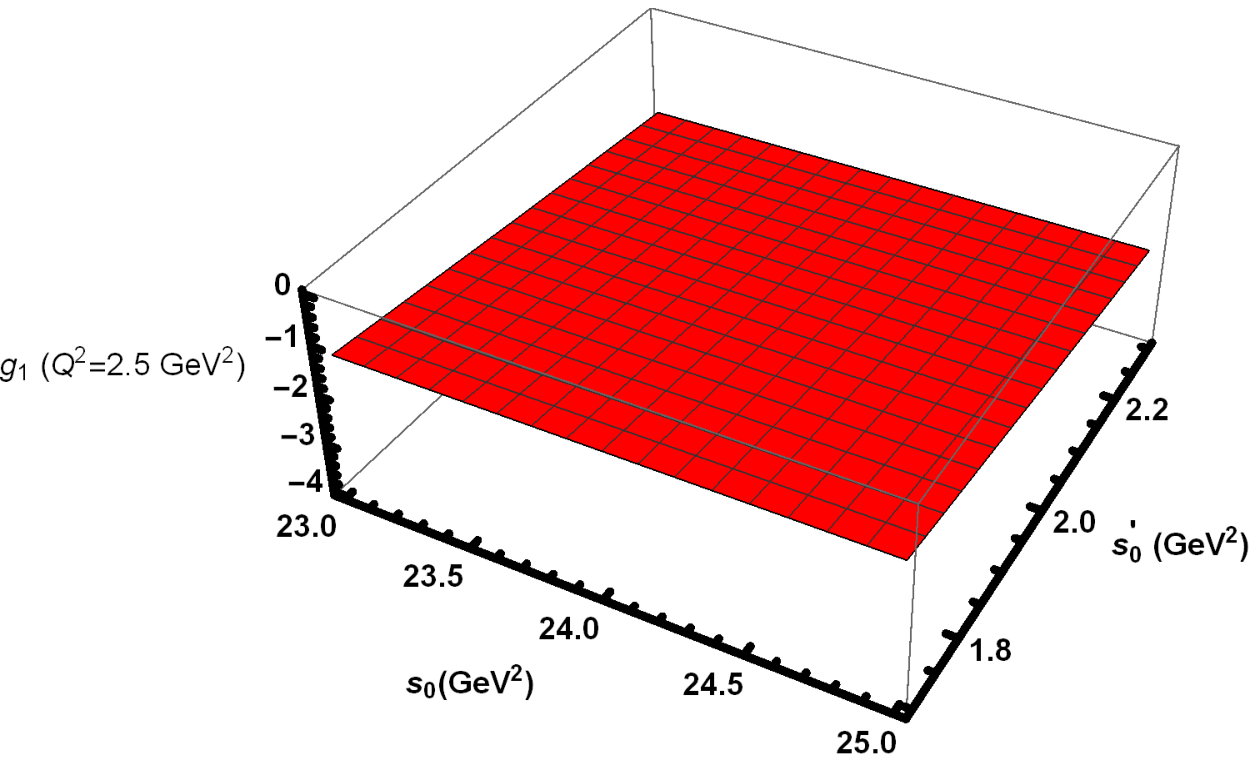}
\end{center}
\caption{\textbf{Left:} Variation of the the coupling constant $g_1$ as function of $M^2$ and $M'^2$ at central values of threshold parameters $s_0$ and $s'_0$ and at $Q^2=2.5~\mathrm{GeV}^2$.
\textbf{Right:} Variation of the the coupling constant $g_1$ as function of $s_0$ and $s'_0$ at central values of Borel parameters $M^2$ and $M'^2$ and at $Q^2=2.5~\mathrm{GeV}^2$. }
\label{gr:g1MsqS0}
\end{figure}

The numerical analyses of the $g_1$ and $g_2$ given in the Eq.~(\ref{eq:g1g2}) require the input parameters given in Table~\ref{tab:Inputs} and some additional auxiliary parameters such as Borel parameters $M^2$, $M'^2$ and threshold parameters $s_0$ and $s'_0$ and the mixing parameter $\beta$ present in the interpolating current of the $\Lambda$ state. The similar standard criteria of the method used for the mass calculation in the previous section, namely weak dependence on the auxiliary parameters, pole dominance, and the convergence of the operator product expansion (OPE) used on the QCD side, are applied in the determination of the auxiliary parameters of this section as well. Taking into account their relations and considering the possible excited resonances of the considered states, the threshold parameters are fixed as:
\begin{eqnarray}
23.0\,\,\mathrm{GeV}^{2}&\leq& s_{0}\leq 25.0\,\,\mathrm{GeV}^{2},
\nonumber \\
1.7\,\,\mathrm{GeV}^{2}&\leq& s'_{0}\leq 2.3\,\,\mathrm{GeV}^{2},
\label{Eq:s0s0p}
\end{eqnarray}
in which the interval of $s_0$ is the same as the one used in the previous section.  At this point, we shall note that the threshold parameter may be expected to be the same, for instance, as in Ref.~\cite{Azizi:2021utt},  given the close masses of the particles involved. However, the analyses performed on the considered state with the given requirements in the previous section are effective in the final determination of these parameters. In each distinct study, it becomes necessary to reassess these requirements. Therefore, in the analyses of the results, one needs to check these requirements in every different work from scratch and obtain the a proper interval satisfying the given requirements for identifying these parameters. Besides, as is seen in Ref.~\cite{Azizi:2021utt}, the structures assigned for these two particles are different. While the structure in Ref.~\cite{Azizi:2021utt} was a diquark-diquark-antiquark, the present work adopts a molecular one. As a result, the discrepancy in threshold values arising from the analyses can also be attributed to these inner quark structures assigned to the considered states.  For the Borel parameters, considering the pole dominance and convergence of the OPE lead us to the following intervals:
\begin{eqnarray}
3.0\,\,\mathrm{GeV}^{2}&\leq& M^2\leq 4.0\,\,\mathrm{GeV}^{2},
\nonumber \\
1.5\,\,\mathrm{GeV}^{2}&\leq& M'^2\leq 2.5\,\,\mathrm{GeV}^{2}.
\label{Eq:s0s0p}
\end{eqnarray}
$M^2$ again spans the same interval given in the previous section. The working interval of the last auxiliary parameter, $\beta$, is determined from a parametric plot of the results given as a function of $\cos \theta$ with $\beta=\tan \theta$ in which the relatively stable regions are considered to fix $\beta$ intervals. These analyses give the following intervals: 
\begin{eqnarray}
-1.0\leq\cos\theta\leq -0.5 ~~~~~\mbox{and} ~~~~~~0.5\leq \cos\theta\leq 1.0. 
\end{eqnarray}  
In all of  these intervals for the auxiliary parameters, we expect weak dependence of the results on these parameters.  To depict this, we provide  the graphs of the strong coupling constant $g_1$ as functions of these auxiliary parameters in Fig.~\ref{gr:g1MsqS0} as  examples: The criteria are satisfied,  dependencies are mild and the uncertainties remain inside the limits allowed by the method.

The analyses give the results reliable only for some regions of the $Q^2$, and therefore to get the coupling constants' values at $Q^2=-m_{J/\psi}^2$, we need to expand the analyses to the region of interest using a proper fit function given as
\begin{eqnarray}
g_i(Q^2)&=& g_{0}e^{c_1\frac{Q^2}{m_{P_{cs}}^2}+c_2(\frac{Q^2}{m_{P_{cs}}^2})^2}.
\end{eqnarray}
The fit parameters providing a good overlap with the results in the reliable region of the QCD sum rule results and the values of the coupling constants obtained from the fit functions at $Q^2=-m_{J/\psi}^2$  are presented in  Table~\ref{tab:FitParam}.
\begin{table}[tbp]
\begin{tabular}{|c|c|c|c|c|}
\hline\hline
Coupling constant&$ g_0$ & $c_1$  & $c_2$ & $g_i(-m_{J/\psi}^2) $ \\ \hline\hline
$g_1$  & $-1.10\pm 0.13$ &$6.43$ & $-26.13$ & $(-4.71 \pm 0.52)\times 10^{-5}$\\ 
$g_2$  &$15.57\pm 1.86$ & $4.43$ & $-3.81$ & $0.61 \pm 0.07$\\
\hline\hline 
Coupling constant&$ g_0$ & $c_1$  & $c_2$ & $g(-m_{\eta_c}^2) $ \\ \hline\hline
$g$  &$7.71\pm 0.85 $ & $5.98$ & $-6.23$ & $ 0.11 \pm  0.02 $\\
\hline\hline 
\end{tabular}%
\caption{ Values of the fit parameters for the fit functions of coupling constants, $g_1$, $g_2$ and $g$  and the coupling constant values at $Q^2=-m_{J/\psi}^2$ and $Q^2=-m_{\eta_c}^2$.} \label{tab:FitParam}
\end{table}
The results contain the errors arising from the uncertainties inherited from both the input parameters and the determinations of the intervals of the auxiliary parameters.

The strong coupling constants determined from the QCD sum rules analyses are applied for the width calculation of the decay $P_{cs}\rightarrow J/\psi \Lambda$, which is performed via the relation 
\begin{eqnarray}
\Gamma &=& \frac{f(m_{P_{cs}},m_{J/\psi},m_{\Lambda}) }{16\pi m_{P_{cs}}^2}\Bigg[-\frac{2 (m_{J/\psi}^2 - (m_{\Lambda} + m_{P_{cs}})^2)}{m_{J/\psi}^2 (m_{\Lambda} + m_{P_{cs}})^2}\Big(g_2^2 m_{J/\psi}^2 (m_{J/\psi}^2 + 2 (m_{\Lambda} - m_{P_{cs}})^2) 
\nonumber\\
&+& 
6 g_1 g_2 m_{J/\psi}^2 (m_{\Lambda} - m_{P_{cs}}) (m_{\Lambda} + m_{P_{cs}}) + 
g_1^2 (2 m_{J/\psi}^2 + (m_{\Lambda} - m_{P_{cs}})^2) (m_{\Lambda} + m_{P_{cs}})^2 \Big)\Bigg].
\label{Eq:DW}
\end{eqnarray}
The function $f(x,y,z)$ in the width formula is given as
\begin{eqnarray}
f(x,y,z)&=&\frac{1}{2x}\sqrt{x^4+y^4+z^4-2xy-2xy-2yz}.
\end{eqnarray}   
The result obtained for the width is
\begin{eqnarray}
\Gamma(P_{cs} \rightarrow J/\psi \Lambda)&=&\left(7.22\pm 1.78\right)~\mathrm{MeV},
 \label{Eq:DWNegativeParity}
\end{eqnarray}
in a nice agreement with the experiment.

\section{QCD sum rule to analyze the $P_{\psi s}^{\Lambda}(4338){}^0\rightarrow \eta_c \Lambda $ decay}\label{IV}

This section provides the investigation of another possible decay channel of the  $P_{\psi s}^{\Lambda}(4338)$ state, namely $P_{\psi s}^{\Lambda}(4338){}^0\rightarrow \eta_c \Lambda $ decay, and the corresponding width for this channel. For the calculation of the corresponding strong coupling constant, the three-point correlation function is
\begin{equation}
\Pi (p, q)=i^2\int d^{4}xe^{-ip\cdot
x}\int d^{4}ye^{ip'\cdot y}\langle 0|\mathcal{T} \{J^{\Lambda}(y)
J^{\eta_c}(0)\bar{J}^{P_{cs}}(x)\}|0\rangle.
\label{eq:CorrF2Pc}
\end{equation}
In addition to the interpolating currents, $ J^{P_{cs}}$ and $ J^{\Lambda} $ previously given, the interpolating current of $\eta_c$ is needed:
\begin{eqnarray}
J^{\eta_c}&=&\bar{c}_l\gamma_{5}c_l.
\label{InterpFieldseta}
\end{eqnarray}
We follow similar steps with the previous section to calculate the width of the strong decay under study in this section.
The hadronic side for this decay is obtained via insertion of complete sets of hadronic states carrying the same quantum numbers of the applied interpolating currents as
\begin{eqnarray}
\Pi^{\mathrm{Had}}(p, q)=\frac{\langle 0|J^{\Lambda }|\Lambda(p',s')\rangle \langle 0|J^{\eta_c}|\eta_c(q)\rangle \langle \eta_c(q) \Lambda(p',s')|P_{cs}(p,s)\rangle \langle P_{cs}(p,s)|\bar{J}^{P_{cs}}|0\rangle }{(m_{\Lambda}^2-p'^2)(m_{\eta_c}^2-q^2)(m_{P_{css}}^2-p^2)}+\cdots,  \label{eq:CorrF1Phys2}
\end{eqnarray}
where $\cdots$ represent the contribution of the higher states and continuum. $p$, $p'$ and $q$ are the  momenta of the  $P_{cs}$,  $\Lambda$ and $\eta_c$ states, respectively. Besides the matrix elements given in Eq.~(\ref{eq:ResPcss}), we need the matrix element relevant to the $\eta_c$ state in terms of the  related decay constant $ f_{\eta_c} $ which is given as
\begin{eqnarray}
\langle 0|J^{\eta_c}|\eta_c(q)\rangle &=&\frac{f_{\eta_c} m_{\eta_c}^2}{2m_c}.
\label{eq:matetac}
\end{eqnarray}
The matrix element, $ \langle \eta_c(q) \Lambda(p',s')|P_{cs}(p,s)\rangle$,  defining this transition is
\begin{eqnarray}
\langle \eta_c(q) \Lambda(p',s')|P_{cs}(p,s)\rangle = g\bar{u}_{\Lambda}(p',s') u_{P_{cs}}(p,s),
\label{eq:coupling2}
\end{eqnarray}
where $g$ is the corresponding strong coupling constant. Using the summation over the spins of the $P_{cs}$ and $\Lambda$ states given in the Eq~(\ref{eq:SumPc}), we get the hadronic side as
\begin{eqnarray}
\tilde{\Pi}^{\mathrm{Had}}(p, q)&=&e^{-\frac{m_{P_{cs}^2}}{M^2}}e^{-\frac{m_{\Lambda}^2}{M'^2}}\frac{g  \lambda_{\Lambda} \lambda_{P_{cs}} f_{\eta_c} m_{\eta_c}^2 }{ 2m_{c}  (m_{\eta_c}^2 + Q^2)}\slashed{p}\slashed{p'}+
\mathrm{other~structures}+\cdots,
\label{eq:haddecay2}
\end{eqnarray}
where  $Q^2=-q^2$ and only the Lorentz structure used in the analyses is presented explicitly, and $\mathrm{other~structures}+\cdots$ denote the contributions coming from excited states and the other present structures.

The QCD side is obtained after plugging in the interpolating currents present in the Eq.~(\ref{eq:CorrF2Pc}) and applying contractions of quark fields using Wick's theorem. The result for this side is obtained as
\begin{eqnarray}
\Pi ^\mathrm{QCD}(p, q)&=&i^2\int d^{4}xe^{-ip\cdot
x}\int d^{4}ye^{ip'\cdot y}\frac{1}{\sqrt{6}} \epsilon^{abc} \epsilon^{a'b'c'}\sum_{i=1}^{2}\Big[\mathrm{Tr}[S_s^{bb'}(y-x)\gamma_5CS_d^T{}^{ca'}(y-x)CA_1^i]A_2^iS_u^{ad'}(y-x)\nonumber\\
 &\times &\gamma_5S_c^{d'l}(x)\gamma_{5}S_c^{lc'}(-x)-2A_2^iS_s^{cb'}(y-x)\gamma_5 C S_d^T{}^{ba'}(y-x)CA_1^iS_u^{ad'}(y-x)\gamma_5S_c^{d'l}(x)\gamma_{5}S_c^{lc'}(-x)\nonumber\\
 &-& A_2^iS_d^{ca'}(y-x)\gamma_5 C S_s^T{}^{bb'}(y-x)CA_1^iS_u^{ad'}(y-x)\gamma_5S_c^{d'l}(x)\gamma_{5}S_c^{lc'}(-x)\Big].
\label{eq:CorrF1QCD2}
\end{eqnarray}
Following the same steps in the previous section, the result obtained for this side for the Lorentz structure $\slashed{p} \slashed{p'}$ is matched with that of  the hadronic side for the same Lorentz structure to get the coupling constant $g$ as
\begin{eqnarray}
g = e^{\frac{m_{P_{cs}^2}}{M^2}}e^{\frac{m_{\Lambda}^2}{M'^2}}\frac{2 m_{c} (m_{\eta_c}^2 + Q^2)}{\lambda_{\Lambda} \lambda_{P_{cs}} f_{\eta_c}  m_{\eta_c}}
\tilde{\Pi},
\label{eq:SRg}
\end{eqnarray}
where we again perform the Borel transformations on the variables $-p'^2$ and $-p^2$, and $\tilde{\Pi}$ represents the Borel transformed $\Pi$  for the QCD side of the calculation corresponding to the mentioned Lorentz structure, $\slashed{p} \slashed{p'}$.

In the analyses of this decay channel, we adopt the input parameters given in Table~\ref{tab:Inputs}, and the same  auxiliary parameters $M^2$, $M'^2$, $s_0$, $s'_0$ and $\beta$ given in the previous section, which works also well for this channel.

To get the value of coupling constant $g$ at $Q^2=-m_{\eta_c}^2$, we again need a proper fit function due to the fact that, as in the Sec.~\ref{III}, the results are only reliable in some regions of the $Q^2$. For this purpose, we use the same form of the fit function given in Sec.~\ref{III} with the fit parameters presented in Table~\ref{tab:FitParam}, providing a good overlap for the obtained results of the QCD sum rule in the reliable region. Note that Table~\ref{tab:FitParam} also contains the numerical value of the coupling constant $g$ at $Q^2=-m_{\eta_c}^2$.

With the obtained coupling constant, the width of the $P_{cs}\rightarrow \eta_c \Lambda$ decay is attained using the following width formula:
\begin{eqnarray}
\Gamma &=& g^2 \frac{f(m_{P_{cs}},m_{\eta_c},m_{\Lambda})}{8\pi m_{P_{cs}}^2}[(m_{\Lambda} + m_{P_{cs}})^2-m_{\eta_c}^2],
\label{Eq:DW2}
\end{eqnarray}
with the function $f(x,y,z)$ given in Sec.~\ref{III}. Finally the width is obtained for this decay channel as
\begin{eqnarray}
\Gamma(P_{cs} \rightarrow \eta_c \Lambda)&=&\left(3.18\pm 0.74\right)~\mathrm{MeV}.
 \label{Eq:DW2}
\end{eqnarray}

\section{Discussion and conclusion}\label{Sum} 

In a recent report, the LHCb collaboration announced the observation of a new candidate pentaquark state with strangeness in $J/\psi \Lambda  $ channel. The observed mass and the width of the state were reported as $m=4338.2\pm 0.7\pm0.4~\mathrm{MeV}$ and $\Gamma=7.0\pm 1.2\pm 1.34~\mathrm{MeV}$, respectively~\cite{LHCb:2022jad} with preferred  spin and  parity quantum numbers being $J^P=\frac{1}{2}^-$. To elucidate its inner structure and certify its quantum numbers, further theoretical investigations are necessary. With this purpose, in the present work, the $P_{\psi s}^{\Lambda}(4338){}^0$ state was  assigned a molecular $\Xi_c\bar{D}$ structure with spin parity $J^P=\frac{1}{2}^-$ and its decays to $J/\psi \Lambda $ and $\eta_c \Lambda $ states were investigated using the three-point QCD sum rule approach. For completeness, firstly, the chosen interpolating current  was applied to calculate the mass and the current coupling constant of the considered state using the two-point QCD sum rule method. These quantities are main  inputs in the decay calculations. The obtained mass, $m_{P_{cs}}=4338\pm 130~\mathrm{MeV}$, is in good consistency with the observed one. Our prediction for the mass is also consistent with the mass predictions based on the molecular assumption present in the literature, such as $m=4327.4$~MeV~\cite{Yang:2022ezl}, $m=4341.0$~MeV~\cite{Ortega:2022uyu}, $m=4336.34$~MeV and $m=4329.11.34$~MeV~\cite{Giachino:2022pws}, and $m=4.34^{+0.07}_{-0.07}$~GeV~\cite{Wang:2022neq}.

As stated, the predicted mass  and the current coupling constant comprise the main input parameters for the width calculations of the  $P_{\psi s}^{\Lambda}(4338){}^0\rightarrow J/\psi \Lambda $ and $P_{\psi s}^{\Lambda}(4338){}^0\rightarrow \eta_c \Lambda $ channels which are taken into account as dominant decay channels of the considered state. To compute the widths of these channels, we first calculated the relevant strong coupling constants and subsequently used them to get the corresponding widths. The resultant widths are obtained as $\Gamma(P_{cs} \rightarrow J/\psi \Lambda)=\left(7.22\pm 1.78\right)~\mathrm{MeV}$ and $\Gamma(P_{cs} \rightarrow \eta_c \Lambda)=\left(3.18\pm 0.74\right)~\mathrm{MeV}$ whose total, $\Gamma =\Gamma(P_{cs} \rightarrow J/\psi \Lambda)+ \Gamma(P_{cs} \rightarrow \eta_c \Lambda)=(10.40\pm 1.93)~\mathrm{MeV}$, also agrees with the experimentally observed width within the presented uncertainties.

Within the existing literature, numerous works have investigated the decays of observed pentaquark states.
In order to elucidate the internal structures of the observed pentaquark states, the width calculations of various transition channels are necessary for either comparing the result with experimental findings or potentially uncovering new channels for further investigations. $J/\psi \Lambda $ is the channel that the observation of $P_{\psi s}^{\Lambda}(4338){}^0$   was reported. However, the width of this particular pentaquark state also receives contributions from other channel, $\eta_c \Lambda$, that was considered in the present study. This is the case for different pentaquark states. If we consider theoretical studies present in the literature on the observed pentaquark states, for instance, with spin-parity quantum numbers $J^P=\frac{1}{2}^-$, as in our case,  the widths were obtained for their decays using different form factor sets in Ref.~\cite{Lin:2019qiv} as $P_c(4312)\rightarrow J/\psi p = 0.001~\mathrm{MeV}$, $P_c(4312)\rightarrow \eta_c p =0.01~\mathrm{MeV}$, $P_c(4312)\rightarrow J/\psi p =0.1~\mathrm{MeV}$, $P_c(4312)\rightarrow \eta_c p =0.4~\mathrm{MeV}$ for $P_c(4312)$ state, $P_c(4440)\rightarrow J/\psi p =0.03~\mathrm{MeV}$, $P_c(4440)\rightarrow \eta_c p =3^{-4}~\mathrm{MeV}$, $P_c(4440)\rightarrow J/\psi p =0.6~\mathrm{MeV}$, $P_c(4440)\rightarrow \eta_c p =0.07~\mathrm{MeV}$ for $P_c(4440)$, and   $P_c(4457)\rightarrow J/\psi p =0.02~\mathrm{MeV}$, $P_c(4457)\rightarrow \eta_c p =2^{-4}~\mathrm{MeV}$, $P_c(4457)\rightarrow J/\psi p =0.2~\mathrm{MeV}$, $P_c(4457)\rightarrow \eta_c p =0.02~\mathrm{MeV}$ for $P_c(4457)$.  In Ref.~\cite{Li:2023zag} the following predictions were obtained: $P_c(4312)\rightarrow J/\psi p = 0.17^{-0.04}_{+0.04}~\mathrm{MeV}$, $P_c(4312)\rightarrow \eta_c p =0.085^{-0.016}_{+0.018}~\mathrm{MeV}$. In  Ref.~\cite{Wang:2019spc} the widths were attained as $P_c(4312)\rightarrow J/\psi p = 0.32\pm 0.08~\mathrm{MeV}$, $P_c(4312)\rightarrow \eta_c p =0.89\pm 0.25~\mathrm{MeV}$, $P_c(4440)\rightarrow J/\psi p = 2.92 \pm 0.55~\mathrm{MeV}$, $P_c(4440)\rightarrow \eta_c p =0.15 \pm 0.03~\mathrm{MeV}$, $P_c(4457)\rightarrow J/\psi p =0.45 \pm 0.13~\mathrm{MeV}$, $P_c(4457)\rightarrow \eta_c p = 0.02 \pm 0.01~\mathrm{MeV}$. In Ref.~\cite{Xu:2019zme} the width predictions were given as $P_c(4312)\rightarrow J/\psi p = 1.67^{+0.92}_{-0.56}~\mathrm{MeV}$, $P_c(4312)\rightarrow \eta_c p =5.54^{+0.75}_{-0.5}~\mathrm{MeV}$. The widths of same transitions were obtained in Ref.~\cite{Dong:2020nwk} as  $P_c(4312)\rightarrow J/\psi p = 0.0448^{+0.0197(+0.0309)}_{-0.0161(-0.0287)}~\mathrm{MeV}$, $P_c(4312)\rightarrow \eta_c p =0.0892^{+0.0392(+0.0615)}_{-0.0321(-0.0571)}~\mathrm{MeV}$. And also the transitions to these final states, for possible spin-parity $J^P=\frac{3}{2}^-$ case, were also considered in Ref.~\cite{Lin:2019qiv} with the following findings:  $P_c(4440)\rightarrow J/\psi p =0.02~\mathrm{MeV}$, $P_c(4440)\rightarrow \eta_c p =8^{-5}~\mathrm{MeV}$, $P_c(4440)\rightarrow J/\psi p =1.8~\mathrm{MeV}$, $P_c(4440)\rightarrow \eta_c p =0.008~\mathrm{MeV}$ for $P_c(4440)$,  $P_c(4457)\rightarrow J/\psi p =0.01~\mathrm{MeV}$, $P_c(4457)\rightarrow \eta_c p =6^{-5}~\mathrm{MeV}$, $P_c(4457)\rightarrow J/\psi p =0.6~\mathrm{MeV}$, $P_c(4457)\rightarrow \eta_c p =0.003~\mathrm{MeV}$ for $P_c(4457)$. As is seen, in some cases, the widths for the channels involving $J/\psi$  in the final state have dominant contributions, while in others, the widths of channels involving $\eta_c$ state are large. The width result obtained for the $P_{\psi s}^{\Lambda}(4338){}^0\rightarrow \eta_c \Lambda $ decay channel in this work indicated that this channel has a significant contribution to the total width, though it is smaller than that of $P_{\psi s}^{\Lambda}(4338){}^0\rightarrow J/\psi \Lambda $   decay channel that the state $P_{\psi s}^{\Lambda}(4338){}^0$   was observed. 

The results obtained for the mass and the total width  of $P_{\psi s}^{\Lambda}(4338){}^0$ are consistent with the experimental findings within the presented uncertainties and favor the $\Xi_c\bar{D}$ molecular nature of the $P_{\psi s}^{\Lambda}(4338){}^0$ state with quantum numbers $J^P=\frac{1}{2}^-$.

\section*{ACKNOWLEDGEMENTS}
K. Azizi is thankful to Iran Science Elites Federation (Saramadan)
for the partial  financial support provided under the grant number ISEF/M/401385.



\end{document}